\documentclass{article}
\usepackage[T1]{fontenc}
\usepackage[utf8]{inputenc}
\usepackage{spconf,amsmath,graphicx}
\usepackage{booktabs}
\usepackage[unicode=true]{hyperref}
\usepackage{gensymb}
\usepackage{hyperref}
\usepackage{caption}
\usepackage{subcaption}

\title{On the relevance of the differences between HRTF\\measurement setups for machine learning}
\twoauthors
 {Johan Pauwels}
	{Queen Mary University of London\\
	j.pauwels@qmul.ac.uk}
 {Lorenzo Picinali}
	{Imperial College London\\
	l.picinali@imperial.ac.uk}
\begin{document}
\ninept
\maketitle
\begin{abstract}
As spatial audio is enjoying a surge in popularity,
data-driven machine learning techniques that have been proven successful in other domains are increasingly used to process head-related transfer function measurements. However, these techniques require much data, whereas the existing datasets are ranging from tens to the low hundreds of datapoints. It therefore becomes attractive to combine multiple of these datasets, although they are measured under different conditions. In this paper, we first establish the common ground between a number of datasets, then we investigate potential pitfalls of mixing datasets. We perform a simple experiment to test the relevance of the remaining differences between datasets when applying machine learning techniques. Finally, we pinpoint the most relevant differences.

\end{abstract}
\section{Introduction}
\label{sec:intro}

The subtle filtering effect caused by the absorption and scattering of sound on the ears, head and torso allow us to perceive the direction of audio sources. Accounting for these effects in the binaural rendering of virtual and augmented sonic environments is therefore necessary in order to achieve maximal realism and immersion. Due to an increasing number of applications in communication, healthcare and entertainment, the technology to create such environments has recently received growing interest from  both academic and industry researchers.

Because of the differences in morphology, the filtering is specific to each individual. A personalised head-related transfer function (HRTF) encodes the direction-dependent filters in the frequency domain, whereas the equivalent representation in the time domain is called a head-related impulse response (HRIR). The acquisition of personalised HRTFs can be done in multiple ways~\cite{guezenoc_hrtf_2018}, of which acoustic measurement and numerical simulation are the most common. These processes all require a dedicated and expensive setup, and are time-intensive, which makes them scale badly to a larger population.


In order to come come up with more scalable ways of obtaining individual HRTFs, several approaches have been designed and evaluated~\cite{picinali2023system}. A promising approach is to use data-drive techniques, for example~\cite{yamamoto_fully_2017, chen_autoencoding_2019, miccini_hrtf_2020} use auto-encoders on HRTFs to discover a latent space that is subsequently used to enable the synthesis of HRTFs in various ways. Alternatively, \cite{chun_deep_2017, fayek_data-driven_2017, wang_global_2021, lee_personalized_2018} synthesise HRTFs from anthropometric measurements, where the last uses pinna images as additional input. Another type of input that is used to generate HRTFs is a 3D model, as examined in \cite{zhou_predictability_2021}. Finally, spatial interpolation of HRTFs is explored in \cite{kestler_head_2019, siripornpitak_spatial_2022}.

Although one motivation behind much of this research is to create procedures that limit the need for the current complex HRTF acquisition methods, the development of these new techniques relies very much on the availability of measured HRTF data. More data typically results in better performance, whereas existing datasets contain only tens to low hundreds of examples. A potential breakthrough could be if we could simply mix data acquired from different measurement setups; whether that is possible or not, and to which extent, is the subject of this paper.

The various measurement setups result in multiple variable elements~\cite{li_measurement_2020}, so first we need to harmonise the data to remove a number of known differences. Then we test the relevance of the remaining differences for machine learning (ML) applications. We postulate that the subjects measured in different measurement setups are all independent samples of the same population. Therefore, in the absence of systematic differences in measurement setups, a classifier should not be able to distinguish different setups to a level higher than chance. 

To test this hypothesis, a classifier will be trained to predict from which dataset an HRTF is taken, i.e.\ under which setup it has been measured. From a practical standpoint, the resulting system has no direct use. There are no HRTFs around for which this information is unknown, but this procedure can teach us whether ML algorithms can potentially exploit dataset-specific characteristics as a shortcut in more realistic ML tasks as well. For example, if an ML algorithm is able to pick up these differences, there is a chance for a machine learning system to internally create distinct sub-models for each setup, which voids the benefits of using multiple datasets to increase training data, and will eventually fail to generalise to unseen datasets. Whether the differences between HRTFs obtained from different setups are perceptually relevant or not is another question, which won't be addressed here.


Apart from the practical implications, also on the conceptual level an HRTF should only contain the filtering effects a specific person has on sound. Undesired characteristics of the setup would otherwise leak into binaural renderings made with that HRTF.


The differences between measurement setups have previously been studied by the ``Club Fritz'' project~\cite{andreopoulou_inter-laboratory_2015}, where the same dummy head was acoustically measured in different laboratories and under different conditions. These measurements were compared also with HRTFs obtained by numerical simulation starting from a geometric model of the head and pinnas \cite{greff_round_2007}. The main difference of our study compared to the Club Fritz project is that we look at multiple human subjects per measurement setup, with no subjects measured under multiple conditions, as opposed to a single dummy head. This introduces additional variance in the measurements, but we argue that this is counteracted by the use of multiple subjects. Only systematic differences are relevant, random variations will be be smoothed out. Furthermore, we are not studying symmetry or inter-aural differences; only single ears are processed and the measurements of right ears are mirrored to reduce variance.


\begin{table*}[ht]
\centering
\begin{tabular}{lrrrrrl}
\toprule 
name & human subjects & samplerate & HRIR duration & radius & \# directions & method\tabularnewline
\midrule
\href{https://www.york.ac.uk/sadie-project/database.html}{SADIE II}
\cite{armstrong_perceptual_2018} & 18 & 96$^*$~kHz & 5.333$^*$ ms & 1.2m & 2114$^*$ & measured\tabularnewline
\href{https://www.akustik.rwth-aachen.de/go/id/lsly/lidx/1}{ITA Aachen}\cite{bomhardt_high-resolution_2016} & 46 & 44.1~kHz & 5.805 ms & 1.2m & 2304 & measured\tabularnewline
\href{https://www.oeaw.ac.at/isf/das-institut/software/hrtf-database}{ARI} & 221 & 48~kHz & 5.333 ms & 1.2m & 1550 & measured\tabularnewline
\href{http://recherche.ircam.fr/equipes/salles/listen/}{Ircam Listen} & 50 & 44.1~kHz & 11.610 ms & 1.95m & 187 & measured\tabularnewline
\href{http://opendap.ircam.fr/SimpleFreeFieldHRIR/BILI}{Ircam BiLi}
\cite{carpentier_measurement_2014} & 55 & 96~kHz & 21.333 ms & 2.06m & 1680 & measured\tabularnewline
\href{https://www.axdesign.co.uk/tools-and-devices/sonicom-hrtf-dataset}{SONICOM} & 120 & 96$^*$~kHz & 5.333$^*$ ms & 1.7m & 864 & measured\tabularnewline
\href{http://www.princeton.edu/3D3A/HRTFMeasurements.html}{Princeton 3D3A} \cite{sridhar_database_2017} & 38 & 96~kHz & 21.333 ms & 0.76m & 648 & measured$^*$\tabularnewline
\href{http://dx.doi.org/10.14279/depositonce-8487}{HUTUBS} \cite{brinkmann_cross-evaluated_2019} & 87 & 44.1~kHz & 5.805 ms & 1.47m & 440 & measured$^*$\tabularnewline
\href{http://sofacoustics.org/data/database/chedar/documentation.pdf}{CHEDAR}
\cite{ghorbal_computed_2020} & 1253 & 48~kHz & 10 ms & 1m$^*$ & 2522$^*$ & simulated\tabularnewline
\href{http://sofacoustics.org/data/database/widespread}{Widespread}
\cite{guezenoc_wide_2020} & 1005 & 48~kHz & 10 ms & 1m$^*$& 2522 & simulated\tabularnewline
\midrule
Common ground & 18 & 41.1~kHz & 5.333 ms & n/a & 12 & n/a\tabularnewline
\bottomrule
\end{tabular}
\caption{\label{tab:datasets}The used HRTF datasets and relevant key properties. An asterisk $^*$ signifies that variants with a different value for that property exist in the dataset, but that this particular one was used.}
\end{table*}

\section{Establishing a common ground for all datasets}
\label{sec:preprocessing}

In order to make the experiment as insightful and fair as possible, it is important to remove known differences between measurement setups as much as possible. Some of these differences will manifest themselves just in the numerical values of the HRTFs, whereas others directly impact the format of the HRTF. Notably, sample rate, HRIR duration and the measurement positions all affect the storage format of the data.

We start by selecting ten datasets of HRTFs, based on their public availability and the presence of shared measurement positions. These datasets are listed in \autoref{tab:datasets}. Eight out of ten are acoustically measured, the remaining two are created with Boundary Element Method simulations from 3D meshes~\cite{ziegelwanger_mesh2hrtf_2015}. In order to avoid any potential spectral distortion associated with interpolating spatial positions~\cite{zhong_maximal_2009}, we only use the intersection of the positions in all datasets, without spatial resampling. For simplicity, we restrict ourselves to the horizontal plane\footnote{Strictly speaking, the HUTUBS dataset does not contain measurements in the horizontal plane, but at an elevation of -0.72\degree. Since all angles are measured up to a considerably larger tolerance, it was deemed close enough.}. The resulting twelve positions are $0\degree, \pm30\degree, \pm60\degree, \pm90\degree, \pm120\degree, \pm150\degree, 180\degree$.

To minimise the differences in loudspeaker and microphone characteristics between recording setups, free-field compensated versions of all HRTFs are used. Not all datasets provide raw recordings and a free-field reference separately, so the versions compensated by the original authors are used. The exact signal processing steps to obtain these is itself a source of variation~\cite{andreopoulou_inter-laboratory_2015}. Therefore minimum-phase versions are calculated to compensate for the differences in onset due to varying windowing parameters.

A common samplerate and HRIR duration is determined by taking the lowest value out of all datasets, respectively 44.1 kHz and 5.333 ms. All HRIRs are resampled and truncated to these values, giving 235 samples long HRIRs.

Because the amplitude of the stimuli and the distance to the loudspeakers differ according to setup, the HRIRs need to be scaled by a dataset-specific factor. Following~\cite{andreopoulou_inter-laboratory_2015}, we take the measurement with the highest RMS (at the ipsilateral position) for both ears of each subject, then calculate the median over an entire dataset. All datasets are then scaled by their respective factor such that the median of all loudest measurement positions is unity.

The resulting HRIRs are then converted to HRTFs using the Fourier transform, and only their magnitude is retained (future developments might be looking also at the phase component). The DC offset is removed and all frequency bins above 18~kHz are discarded, again following~\cite{andreopoulou_inter-laboratory_2015}.

The final data representation are matrices of 12 angles by 95 frequency bins for each ear of every subject in all datasets, which we believe to be harmonised as well as possible given the available HRTF formats. This will serve as input to the classifier, and its target class will be the name of the respective dataset.


\section{Classification of HRTFs by measurement setup}

\begin{figure}[htb]
    \centering
    \begin{subfigure}[b]{\linewidth}
         \centering
         \includegraphics[width=1\linewidth]{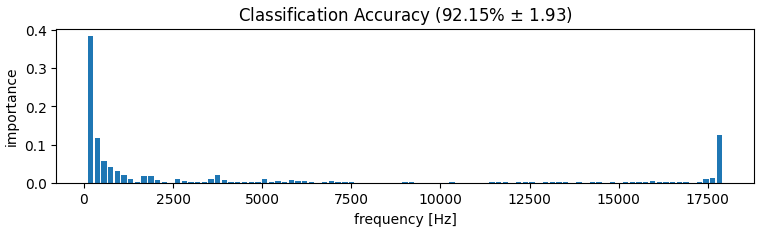}
         \vspace{-6mm}
         \caption{\footnotesize frequency range 187~Hz--18kHz}
         \label{fig:importance-20-18k}
    \end{subfigure}
    \begin{subfigure}[b]{\linewidth}
         \centering
         \includegraphics[width=\linewidth]{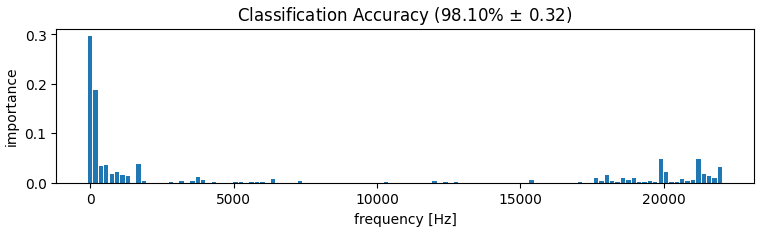}
         \vspace{-6mm}
         \caption{\footnotesize frequency range 0~Hz--22kHz}
         \label{fig:importance-0-22k}
    \end{subfigure}
    \begin{subfigure}[b]{\linewidth}
         \centering
         \includegraphics[width=\linewidth]{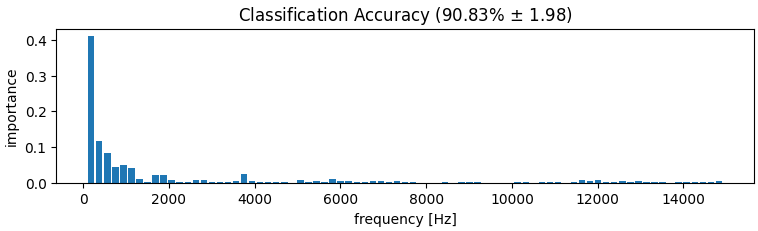}
         \vspace{-6mm}
         \caption{\footnotesize frequency range 187~Hz--15kHz}
         \label{fig:importance-20-15k}
    \end{subfigure}
    \begin{subfigure}[b]{\linewidth}
         \centering
         \includegraphics[width=\linewidth]{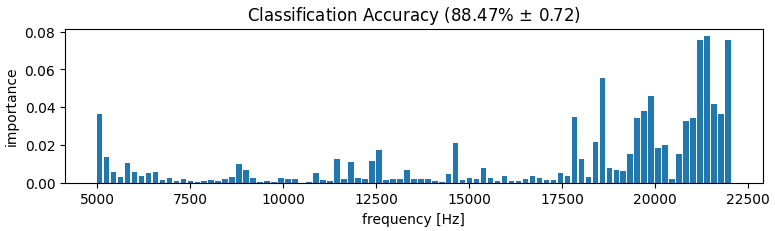}
         \vspace{-6mm}
         \caption{\footnotesize frequency range 5~kHz--22kHz}
         \label{fig:importance-5k-22k}
    \end{subfigure}
    \begin{subfigure}[b]{\linewidth}
         \centering
         \includegraphics[width=\linewidth]{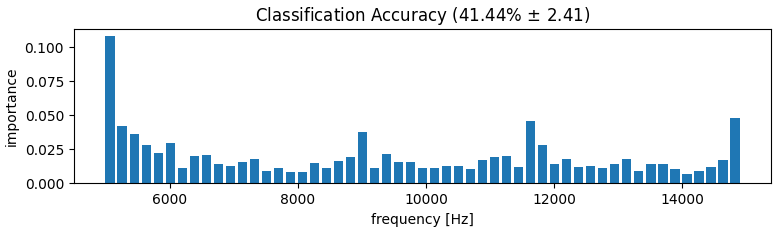}
         \vspace{-6mm}
         \caption{\footnotesize frequency range 5~kHz--15kHz}
         \label{fig:importance-5k-15k}
    \end{subfigure}
    \caption{Frequency importance and five-fold cross validation test accuracy using gradient boosted trees}
    \label{fig:frequency-importance}
\end{figure}

As mentioned before, the measurements taken at the right ear are mirrored around the median plane such that they can be considered as additional examples of left ear measurements, again to simplify the classification problem. For each dataset, we therefore have twice the number of datapoints as the number of subjects listed in \autoref{tab:datasets}. As can be seen in the table, the number of points per dataset differ by orders of magnitude, which is challenging when training a classifier. Therefore we create a balanced dataset by taking the first 36 datapoints of each class, which is the minimum as determined by the SADIE II dataset.

Due to the spatial symmetry between ears of the same subject~\cite{zhong_spatial_2013}, the left and the mirrored-right datapoints can obviously not be considered as independent. Consequently, they are treated as indivisible groups when splitting the dataset into training and test sets.

The effectiveness of the classifier is measured using its accuracy on the test set, i.e.\ the percentage of unseen examples that are recognised correctly. While every effort is made to achieve an accuracy that is as high as possible, in the ideal scenario this would be limited to the level of random guessing (10\%). That would mean that the preprocessing, as detailed in \autoref{sec:preprocessing}, is enough to make measurement setups indistinguishable from each other, and therefore that datasets can be combined without drawbacks.

Perhaps surprisingly, our results show that the average testing accuracy over 5 cross-validation folds can achieve more than 95\%  with relatively simple CART decision tree~\cite{breiman_classification_2017} and linear support vector machine (SVM)~\cite{fan_liblinear_2008} classifiers. The average training accuracy reaches 100\%, indicating some overfitting. The few incorrectly recognised examples are distributed over multiple datasets, without strong trends. We can conclude that each dataset has a pattern of frequencies observable in one or a combination of spatial positions that sets it apart from other datasets regardless of the specific subject, i.e.\ it acts as a fingerprint that can be used to identify the measurement setup.

Originally, these experiments were performed using deep learning based classifiers, but these were substituted for simpler classical machine learning algorithms once it was confirmed that these findings could be replicated with classical ML classifiers. Not only are the latter faster to train and easier to interpret, they also allow to draw stronger conclusions from these results. These standard classifiers operate directly on the features, in our case the HRTF spectra, meaning that the identifying patterns are formed by combining frequencies directly. In contrast, the premise of deep learning is that first optimal features get derived from the inputs before those features are combined to predict the output.

\begin{figure*}[htb]
    \centering
    \begin{subfigure}[b]{0.95\columnwidth}
         \centering
         \includegraphics[width=1\linewidth]{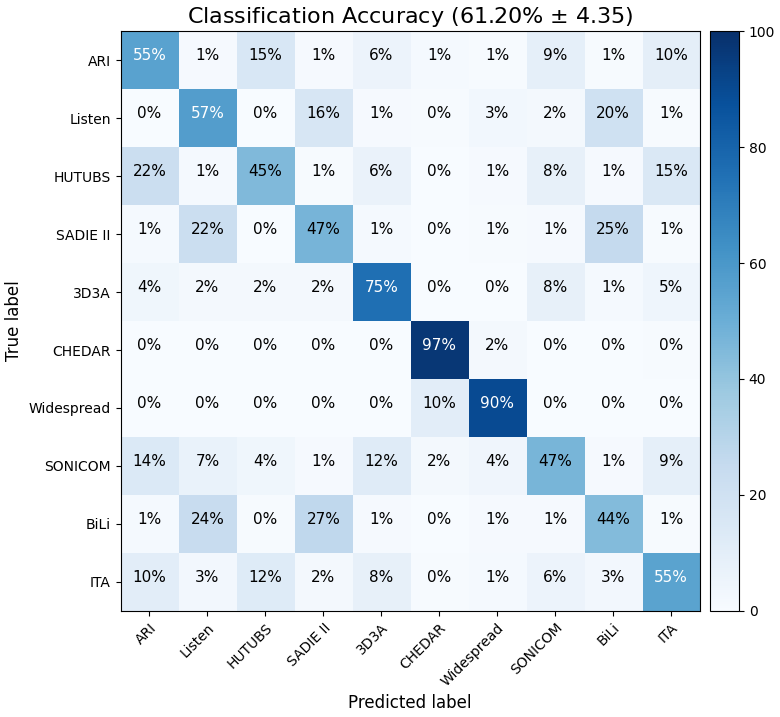}
         \vspace{-6mm}
         \caption{Radial basis function SVM classifier}
         \label{fig:conf-svm}
    \end{subfigure}
	\hfill
    \begin{subfigure}[b]{0.95\columnwidth}
         \centering
         \includegraphics[width=1\linewidth]{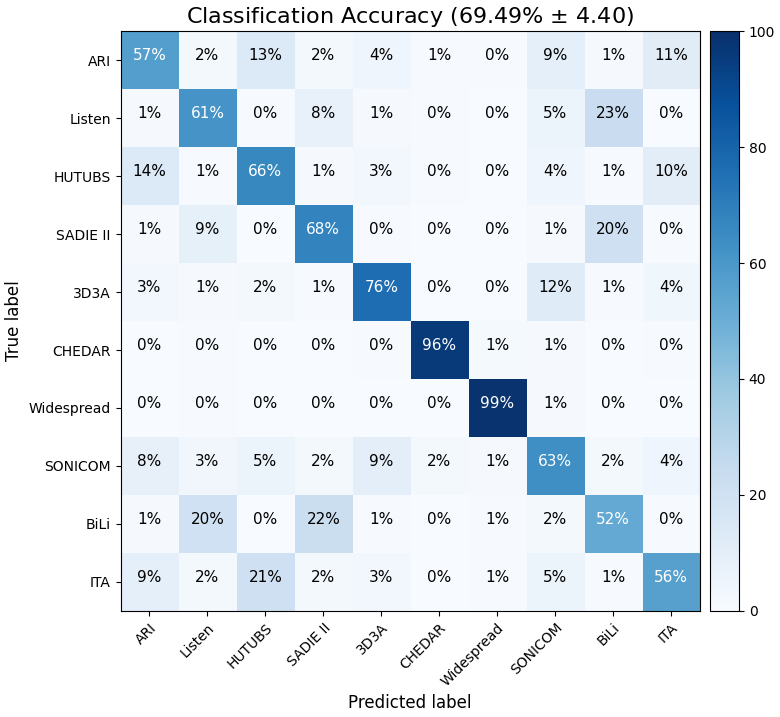}
         \vspace{-6mm}
         \caption{Gradient boosted trees classifier}
         \label{fig:conf-gbt}
    \end{subfigure}
    
    \caption{Confusion matrices for the classification of individual source position in HRTFs limited to frequency range 1--18~kHz}
    \label{fig:confusion-matrices}
\end{figure*}

\section{Classification of individual HRTF positions by measurement setup}

The previous experiment demonstrated that, when all source positions are observed at once, at least one reveals the measurement setup. This excellent classification accuracy could be due to a single or a combination/contrasting of multiple positions, which might also differ between specific datasets. In order to study the direction-independence of the differences between measurement setups, another classification experiment is performed.

The multi-directional HRTFs are broken down into individual datapoints and the measurement angle information is discarded, while the target class remains the name of the dataset. The number of datapoints per class therefore increases by twelve, from 36 to 432. This classification is expected to be harder, since it requires finding identifying frequency patterns for each source position.

Nevertheless, a decision tree trained with 5-fold cross-validation still recognises the dataset with around 81\% accuracy on average and a linear SVM achieves $\pm$ 83\%. The accuracy rises further when slightly more complicated classifiers~\cite{hastie_elements_2009} are used, over 86\% for a non-linear SVM with a radial basis function kernel and more than 92\% for gradient-boosted trees. In addition to being the highest performing classifier, gradient boosted trees have the advantage that they are easy to interpret and therefore can provide the relative importance of each frequency in the decision process, visualised in \autoref{fig:importance-20-18k}. Because we removed the DC offset as described in~\autoref{sec:preprocessing}, the lowest remaining frequency is $\frac{44100}{235} \approx 187~\mathrm{Hz}$. The highest remaining frequency after harmonising the datasets is 18~kHz.

It is immediately clear that the lowest and highest frequencies are contributing the most to the classification process. Note that this importance is not correlated to signal strength. The low frequencies contain most of the signal's energy, and it is likely that the room reflections are being identified as a fingerprint of the measurement setup. In contrast, the high frequencies contain little energy, so the high-frequency ``noise'' is not masked and is consistent enough to contribute to the fingerprint.

To further investigate the importance of those particular frequency ranges, we change the harmonisation procedure slightly and repeat the experiments with different frequency ranges for the HRTFs. Using the full frequency range up to 22~kHz, including DC offset, makes it even easier to identify the dataset, as seen in \autoref{fig:importance-0-22k}.
Excluding more high frequencies, which are inaudible for a large part of the adult population anyway, has relatively little effect, as seen for example in \autoref{fig:importance-20-15k} where all frequencies above 15~kHz are dropped. The classifier seems to partially compensate by using the low frequency info more effectively. The reverse is also true, to a lesser extent. Removing the lower frequencies (below 5~kHz in \autoref{fig:importance-5k-22k}) makes the classifier turn to the highest frequencies to regain most of the classification accuracy. This action is far more destructive for an HRTF though. Only when both the high and low frequencies are removed does the accuracy drop significantly, though it remains significantly above chance level (about 42\% for a frequency range of 5 to 15~kHz, visualised in \autoref{fig:importance-5k-15k}).

\section{Discovering similar setups through confusion pairs}

Imperfect classifiers reveal which classes are the hardest to separate through their confusion matrix. Although such analysis is inherently linked to the type of classifier (different types might struggle with different pairs) without guarantee of generalisation, it is nonetheless insightful to have a look at some confusion matrices from the previous experiment. In \autoref{fig:confusion-matrices}, the confusion for a non-linear SVM classifier and a gradient boosted trees classifier is shown. The range of the HRTF input was limited to 1--18~kHz, such that a fair amount of errors was made, and all source positions are processed individually.

Some interesting observations are common to both classifiers. The CHEDAR and Widespread datasets are particularly easy to distinguish, likely because they both are generated as numerical simulations from morphable models of pinna, head and torso for the former and only pinna for the latter. Furthermore, the Listen, BiLi and SADIE II form a group of datasets that is harder to distinguish. The first two were measured at the same institution, so even though the measurement procedure changed somewhat in between them, such similarity can be expected. The confusion with the SADIE II dataset is harder to explain, but could be due to similar postprocessing of the measurements or sheer luck. ARI, ITA and HUTUBS form a similar group without obvious explanation.

\section{Conclusion and future work}

In this paper, we demonstrated that machine learning classifiers can easily identify the measurement setup in HRTFs, for all employed source positions, and despite preprocessing aimed at harmonising datasets as well as possible. The implication is that any machine learning workflow that involves HRTFs is at risk of generalising badly to other measurement setups, and therefore that cross-dataset testing is of paramount importance. It also means that simply mixing (harmonised) datasets to increase the number of training examples will not automatically lead to an increase in robustness.

It would be interesting to pinpoint the unique identifying characteristics for each dataset in the future, possibly looking also at phase-related differences. This could be done by training binary classifiers between a single dataset and all others. Furthermore, the perceptual relevance of these differences should be studied.

Ideally, the differences between measurement setups are not just diagnosed, but also reduced. It might be possible to do so manually through specific compensation filters once the exact differences are pinpointed. Alternatively, since the experimental procedure proposed in this paper is effectively a discriminator that should be fooled, it could be possible to use adversarial training to create a system that learns how to perform such equalisation automatically. In any case, perceptual evaluation of such compensation and/or other modifications would be required to ensure that the HRTFs keep their relevant characteristics. After all, flat spectra are impossible to distinguish, but also useless.

\section{Acknowledgements}
This research has been partly funded by the SONICOM project (EU Horizon 2020 RIA grant agreement ID: 101017743).



\end{document}